# Mapping of the Influenza-A Hemagglutinin Serotypes Evolution by the ISSCOR Method

Jan P. Radomski [1,4*], Piotr P. Słonimski [2γ], Włodzimierz Zagórski-Ostoja[3], and Piotr Borowicz[4]

[1] *Interdisciplinary Center for Mathematical and Computational Modeling, Warsaw University, Pawińskiego 5A, Bldg. D, PL–02106 Warsaw, Poland*
[2] *Centre de Génétique Moléculaire du CNRS & Université Pierre-et-Marie Curie (Paris-6), 91190 Gif–sur–Yvette, France*
[3] *Institute of Biochemistry and Biophysics, Polish Academy of Sciences, Pawińskiego 5A, Bldg. D, PL–02106 Warsaw, Poland*
[4] *Institute of Biotechnology and Antibiotics, Starościńska 5, PL–02516 Warsaw, Poland*

## *Abstract*

Analyses and visualizations by the ISSCOR method of influenza virus hemagglutinin genes of different A-subtypes revealed some rather striking temporal relationships between groups of individual gene subsets. Based on these findings we consider application of the ISSCOR-PCA method for analyses of large sets of homologous genes to be a worthwhile addition to a toolbox of genomics – allowing for a rapid diagnostics of trends, and ultimately even aiding an early warning of newly emerging epidemiological threats.

## *1. Introduction*

Living organisms have very often quite biased preferences for synonymous encoding of the same amino acids. These differences and their variation have been extensively studied, however, no decisive governing rules have yet been discovered. Frequencies of codons for many species are in close correlation with their genome's GC contents, but the underlying forces governing this are not clear – it might be possible, that it is the GC content which is determining a genome's amino acids predilection for the specific codons being used and their bias [1]. On the other hand, it might be that reverse causative relationships are in operation: codons-specific amino acids usage is a driving factor for observed GC contents. Possible factors and forces driving synonymous codons usage postulated so far include, among many others: translational optimization [2-6], mRNA structural effects [7], protein composition [8], and protein structure [9], gene expression levels [2, 10], the tRNA abundance differences between different genomes, and tRNA optimization [11-13], different mutation rates and patterns [14]. Also, some other possibilities were hypothesized, like local compositional bias [15], and even gene lengths might play a role too [16].

It is clear, that many interesting biological mechanisms underlie the basic phenomenon of genetic code degeneracy. One of its aspects, however, has not been studied until recently [17] – the question dealing with the sequential order of occurrence of synonymous codons. Obviously, an order of elements in a linear set is a different property, than the frequency of elements in the set. The amino acid composition of a protein carries much less information than the amino acid sequence of such protein, which in turn is less information intensive than a corresponding nucleotide sequence coding the same protein. This question can be formulated more precisely if we consider a given frequency of synonymous codon usage characteristic for a gene. There is a very large number of different orders in which the synonymous codons can appear sequentially along the gene without changing either the amino acid sequence of the encoded protein, or the codon usage of the gene.

---

[*] corresponding author: janr@icm.edu.pl
[γ] deceased on April 25th, 2009

Influenza viruses are antigenically variable pathogens, capable to continuously evading immune response. Influenza epidemics in humans cause an estimated 500,000 deaths worldwide per year. The genome of influenza A viruses consists of eight RNA segments that code for 10 viral proteins. Based on the antigenic specificities of the hemagglutinin (HA), or neuraminidase (NA) proteins the influenza A viruses have been divided respectively into 16 HA (H1-H16), and nine neuraminidase (N1-N9) subtypes. Accumulation of mutations in the antigenic sites of the HA and NA, altering viral antigenicity, is called the ''antigenic drift''. In circulating influenza viruses this antigenic drift is a major process, accumulating mutations at the antibody binding sites of receptor proteins, and enabling the virus to evade recognition by hosts' antibodies. The HA protein consists of two domains, HA-1 and HA-2 – the HA-1 domain, the major antigenic protein of influenza A viruses, contains all of the antigenic sites of HA, and it is under constant immune-driven selection. The segmented nature of influenza genome allows also for exchange of gene segments – a process of genetic reassortment, involving type A influenza viruses of different subtypes, and may result in the so called ''antigenic shift'', which occurs when progeny viruses that possess a novel HA, or a novel HA and NA, emerge [18], [19].

Recently we have proposed an *in silico* method [17] to tackle the problem of the sequential order of synonymous codons, called ISSCOR (Intragenic, Stochastic Synonymous Codon Occurrence Replacement) – synonymous codons, which occur at different positions of an ORF are replaced randomly by a Monte Carlo routine with their equivalents – the method generates nucleotide sequences of non-original ORFs, which have identical codon usages, and would encode identical amino acid sequences. The ISSCOR method was then used to analyze temporal and spatial aspects of the three sets of orthologous gene sequences isolated from various strains of hemagglutinin of the influenza A virus subtypes: A/H3N2, A/H1N1 (of both the seasonal, and the 2009 pandemic variants), and A/H5N1 [20] in an alignment-free manner.

The rich collection of the data gathered recently during the last bouts of the pandemic H1N1 flu gave possibility to give a fresh look on the perennial questions of influenza epidemiology. The role of founder effects is important for epidemiological scenarios [21, 22] assuming that a genetic variability common to a small founder population will then also be found in most descendants. In viral outbreaks, such effects can be at play when specific mutations are enriched in samples coming from the same region, and/or the same time. Considering phylogenetic relations it is useful to identify such viral lineage founder events. The global strain sequencing efforts, combined with robust statistics allow novel insights into phylogeny, and especially variability of this highly changeable RNA virus. The variability problem is of especially high interest in view of recent results concerning the switching of receptor selection by the hemagglutinin [23, 24] leading to a possible acquiring of airborne infection transmissibility for mammalian hosts. Russell *et al.* [25], based on the combined results of Imai *et al.* [23] and Herfst *et al.* [24], proposed a mathematical model of within-host H5N1 virus evolution to study some aspects influencing increase or decrease in probability of subsequent substitutions leading to aforementioned switch. They stressed that more data are needed for assessing calculated evolution rates based on the assumed mutation rates, which are of high interest for weighting out the speed of evolution of HA in switching receptor selection. In the current work we have postulated that rates of mutation frequencies in HA commonly accepted are routinely overestimated for at least one order, and proposed an enhanced method of finding evolutionary correlations between multiple strains of the H1N1 2009 pandemic virus [26].

The goal of the current work is to explore in some detail relationships between different orthologous hemagglutinin sequences of various influenza-A serotypes, using their ISSCOR descriptors. As these descriptors are easy to calculate, and yet encompass a rich spatial information involving also long range interactions, together with immediate neighborhoods of constituting residues, it should be possible to correlate stochastic genes' representation based upon the deviates, and their possible 3D epitopic functional interactions – either through theoretical 3D modeling, or perhaps through building associations with appropriate (*in vitro?*) *in vivo* data.

## 2. Materials and Methods

The A-influenza hemagglutinin full length gene sequence data, isolated mostly from avian, human and swine hosts, available for serotypes: H1N1 (seasonal), H1N1 (2009-2010 pandemic), H1N2, H2Nx (all neuraminidases), H3N2, H5N1, and H7Nx (all neuraminidases) were obtained on Dec. 9$^{th}$ 2012 from the NCBI influenza resource. From this collection unique sequences were selected – such that from each subset of identical genes, the ones with the earliest dates of sample isolation were chosen as representatives. **Table I** shows the distribution of the number of sequences for each serotype per hosts.

| serotype | host | sequences |
|---|---|---|
| H1N1s | avian | 116 |
| H1N1s | human | 1280 |
| H1N1s | swine | 678 |
| H1N1v | avian | 4 |
| H1N1v | human | 3243 |
| H1N1v | swine | 197 |
| H1N2 | avian | 10 |
| H1N2 | human | 21 |
| H1N2 | swine | 58 |
| H2Nx | avian | 108 |
| H2Nx | human | 5 |
| H3N2 | avian | 98 |
| H3N2 | human | 2356 |
| H3N2 | swine | 292 |
| H5N1 | avian | 539 |
| H5N1 | human | 107 |
| H5N1 | swine | 17 |
| H5N1 | ferret | 2 |
| H7Nx | all | 88 |
| **All** | **all** | **9131** |

**Table I**

*Overview of the ISSCOR approach:*

Previously [27, 28], we have described alignment free approaches to the problem of comparison and analysis of complete genomes, and some techniques enabling to cope with the sparseness of the n-gram type of genomic information representations. The problem of sparse occurrence matrices is not only present, but even more pronounced when dealing with the number of permutations of the possible synonymous codons. Calculating the set of n–grams for such occurrences will lead to a vector representations, which are severely sparse, especially for higher n–grams lengths, and hence to very poor statistics. To alleviate this problem, we proposed [17] a hybrid approach. Namely, when computing counts of codon-pair patterns – separated by codon sub–sequences of differing length – the actual composition of these spacer sub–sequences will be neglected. However, when such partial counts are used as a composite set, poor statistics is no longer a hindering obstacle, and the complete information about particular n–gram frequencies profile is preserved, albeit in a distributed and convoluted form.

For every protein coding gene, with its original nucleotide sequence $j_0$, a set of equivalent nucleotide strings ($j_1, j_2, j_3,..., j_N$) is created by a Monte Carlo approach. These artificial sequences have the following properties:
- they are all of the same nucleotide lengths as the $j_0$;
- they have exactly the same amino acid sequence as the $j_0$ (i.e., the proteins translated from the $j_1, j_2, j_3,..., j_N$ are identical to $j_0$);
- they have in the vast majority of cases a synonymous codon order <u>different from the original sequence $j_0$.</u>

The last is an essential point, which merits a commentary. The probability that a given string $j_i$ generated stochastically has the same synonymous codon order as the original $j_0$ decreases with the product of its length, with a probability limit tending rapidly to zero.

Therefore, the ISSCOR method allows comparing the original codon sequence with an ensemble of different synonymous sequences – yet all of them coding for the same sequence of amino acids.

*The Computational Procedure:*

Full description of the method is given in [17], however, mathematical steps are briefly outlined here for convenience. First, the codon usage frequencies are determined, and on that basis the probabilities of replacement are calculated, separately for each codon-degeneracy equivalence group $E$:

$$P^k = \frac{u_k}{\sum_{d=1}^{E} u_d} \qquad (1)$$

where:

the $P^k$ – probability that any other codon from the same degeneracy equivalence group $E$ – will be randomly replaced by the codon $k$; and $u_k$ is the synonymous codon $k$ triplet frequency for a given amino acid in a whole gene. Therefore, obviously for any given degeneracy equivalence group $E$, the sum of such probabilities will always be equal to 1.

Then, successively for each codon in a gene the procedure of it's synonymous random replacement is performed based on probabilities according to the equation (**1**). Finally, the resulting shuffled sequences are determined, and compared to the original sequence of the gene.

For each protein coding sequence, we need to determine a complete matrix of all codon-pair patterns. Obviously, in a protein coding sequence, there are at most 3904 (61x61 + 61x3) unique codon-pair patterns. For a given sequence *V*, and the all codon-spacer lengths, in order to calculate observed values of a particular codon-pair pattern (*$c_k$*, *$c_l$*), first we need to construct a series of matrices $O^\lambda$ (*occurrence matrices*). Each element of every matrix $O^\lambda$ contains the counted sum of all specific codon-pair patterns (*$c_k$*, *$c_l$*), separated by a string of other codons present in this sequence, where the $\lambda$ denotes the number of other codons separating the given codon-pair pattern (*$c_k$*, *$c_l$*). Using a sliding window of the length 3*($\lambda$+2) nucleotides, and starting at the position *m*, we would scan the whole sequence *V*, calculating elements of the matrix by the formula:

$$O^\lambda(c_k, c_l, p) = \sum_{m=1}^{M-\lambda-1} f(c_k, c_l, \lambda, m, p), \qquad (2)$$

where M is the sequence's length, and

$$f(c_k, c_l, \lambda, m, p) = \begin{cases} \textbf{1,} \text{ if } V(m) = c_k, \text{ and } V(m+\lambda+1) = c_l, \text{ and the} \\ \quad \text{codon-pair pattern } (c_k, c_l) \text{ matches the pattern of} \\ \quad \text{the particular comparison } p \\ \textbf{0,} \text{ otherwise.} \end{cases}$$

Comparisons involve matches between the predefined codon-pair patterns, of the first codon *$c_k$*, always taken together with the second codon *$c_l$*. That is, a particular positional comparison *p* involves only one nucleotide from the first codon *$c_k$*, and one nucleotide from the second codon *$c_l$*, ignoring all four remaining nucleotides, which corresponds to a pattern (for convenience we name each such pattern a *hexon*). Thus, there are e.g., nine patterns containing the adenine (**A**) at any position in a first codon, together with the cytosine (**C**) at any position in a second codon, etc. Obviously, when $\lambda$ = 0, one has an adjacent codon-pair pattern (hexanucleotide), for $\lambda$ = 1 it is a nonanucleotide, and so on. Note, that since these are ordered counts, each starting at the sequence's 5'–terminus, the matrices $O_i^\lambda$ are not symmetrical, that is the count of the pair (*$c_k$*, *$c_l$*) is different from the count of the pair (*$c_j$*, *$c_k$*).

To make the results independent of a particular sequence size (or a set of sequences, as described already in [17] on the example of the complete genome of *Helicobacter pylori*), we need to calculate how much the number of actually observed *hexons* in the original sequence, differs from the mean number of the corresponding *hexons*, observed after performing *N* number of random ISSCOR permutations, divided by the standard deviation observed in the corresponding shuffled sequences:

$$D_{xAx\_\lambda\_Txx} = \frac{O_{occurences} - \frac{\sum_{n}^{N} R^{n}_{shuffled}}{N}}{STD_{shuffled}} \quad (3)$$

where:

$D_{xAx\_\lambda\_Txx}$ is a **deviate** of the results for, e.g., the pattern *xAx_λ_Txx*, that is for the all codon combinations comprising the nucleotide *A* at the second position in the first codon, and the nucleotide *T* at the first position of the second codon – the border codons being separated by the number $\lambda$ of other codons;

$O_{occurences}$ are the numbers of the actually observed occurrences for any given hexon in the unperturbed sequence;

$R^{n}_{shuffled}$ are the numbers of occurrences for any given hexon pattern counted **after** codons of the whole sequence have been shuffled randomly (as described above), thus the $R^{n}_{shuffled}/N$ is a mean number of such occurrences after the N such random shuffles;

$STD_{shuffled}$ is a standard deviation for all occurrences of a given hexon pattern, after *N* random shufflings of the whole sequence (we have determined previously [17] that 500 shuffles are sufficient to obtain systematic and highly repetitive results).

Phylogenetic analyses were performed using the classic Neighbor Joining [29, 30], and the modified NJ algorithm: QPF [31]. Also the results from the II-iteration trees after multiple alignment, available through the MUSCLE package [32], were checked for consistency with distance-based methods. For tree manipulations (in the Newick format) and their visualization the Dendroscope package of Huson *et al.* was used [33].

## *3. Results and Discussion*

The ISSCOR deviates (equation 3) for all the 9131 hemagglutinin sequences collected were calculated as described earlier, using codon spacer values from λ=0 to λ=16, and creating the matrix **MA** of 9131 rows by 2448 columns. The results of principal component analysis (PCA) for the matrix MA showing 9131 data points (marked in light gray), will be presented on a series of figures (**Figs. 1** to **4**), showing maps of PC-1 values (45.4% of a total variance explained) – plotted on the abscissas, and the PC-2 values (further 18.3% of a total variance) – plotted on the ordinate axes respectively. *The intent is that the 9131 points present in all the maps will provide a common frame of reference*. On such a background the points corresponding to the different serotypic subsets of genes will be color-coded, and grouped also by the hosts infected (avian, human, swine, or in few cases also ferrets).

### *Chronology of evolution*

The presumed trend in a chain of infectivity from avian, through porcine, to human hosts can be well observed e.g. in the case of H3N2 serotype on the right hand column panels in **Figs. 2A** and **2B**. However, it is interesting that for the 2009 pandemic H1N1 strains, isolated from avian hosts, there was most probably a reversal of influenza infectivity chain – although it is usually considered that the avian hosts viruses form a primeval reservoir of infections; and the process of the propagation follows from avian, through mammals (mostly porcine) to human hosts. Such a conclusion, stemming purely just from the analyses of the maps here, was subsequently confirmed by the search in original literature**:** there are only four such isolates present in the NCBI database to date (accessions: HM370960, HM370967, HM370975, HM450134) – all quite late, isolated in Canada from turkeys [34] between Oct. and Dec. 2009. These sequences group on the map together with the cluster of the isolates from human and swine hosts (the left column on **Fig. 1**).

In contrast to the H3N2 strains, which all form together one large, elongated cluster, with a clearly visible evolutionary trend (right side of **Figs. 2A** and **2B;** *c.f.* also [20]), the behavior of all other isolates (H1N1, H5N1, H1N2, H2Nx, and H7Nx) is much more complex, as they intermingle, forming a network of possible evolutionary paths. It is possible to roughly trace an early chronology of e.g. human and porcine of the seasonal H1N1 isolates (**Fig. 1**) from the earliest complete gene sequence of 1918, and then successively from 1930', 40', 50', and so on. All the 9131 points do form an approximately triangular shape on each map, with the most recent strains of the 2009 H1N1 pandemic occupying the leftmost, bottom apex, the most recent human isolates of the H3N2 the rightmost, bottom one, and the topmost position by the human seasonal H1N1 strains of the I-st decade of XXI c. (thus the H1N1 strains are forming the left elongated, crescent-shaped edge; however, with some avian and porcine isolates already stretching towards regions occupied by the H3N2 ones).

Assuming that a hypothetical, primeval ancient influenza hemagglutinin might have occupied a position somewhere in the middle of this triangle, we can follow the chronological spreading of more recent strains towards more and more remote locations (*c.f.* **Fig. 1**). Therefore, it seems reasonable to assume that HA orthologs occupying positions in apexes of this triangular distribution can be considered to mark the extents of this gene changeability in influenza A-type viruses, although of course further genetic drift might expand these boundaries even more. The HA sequences of the three most extended positions on the map are all obtained from isolates of human hosts**:** the topmost – A/Hamburg/1/2005 (H1N1 seasonal; FJ231765), bottom-left – A/Sydney/DD3_17/2010 (H1N1 pandemic, CY092550); and the bottom-right – A/Thailand/Siriraj-06/2002 (H3N2, JN617982), These observations are consistent with the hypothesis that all influenza viruses originated from their ancient stock harbored in wild, migratory aquatic birds – accordingly the avian host isolates are, on average, closest to the center of the triangular spread of HA genes observed here, followed by porcine isolates, and only then isolates from humans.

*Putative origins of the XX c. pandemics*

The question arises then whether it might be possible to trace back an origin of genetic shifts in HA leading to known major pandemic outbreaks. There is clearly a complete dearth of data of sequences prior to the H1N1 1918 strain. Also, for the next major pandemic emergence in 1957 the H2N2 viruses that caused the Asian flu, but disappeared from human population a decade later. There are only six complete HA H2N2 genes present in this set, which means not sufficient information to draw valid conclusions. The **Table II** contains the distance matrix, showing the respective numbers of nucleotide differences between these strains (*c.f.* also **Fig. 3B**, top-right panel).

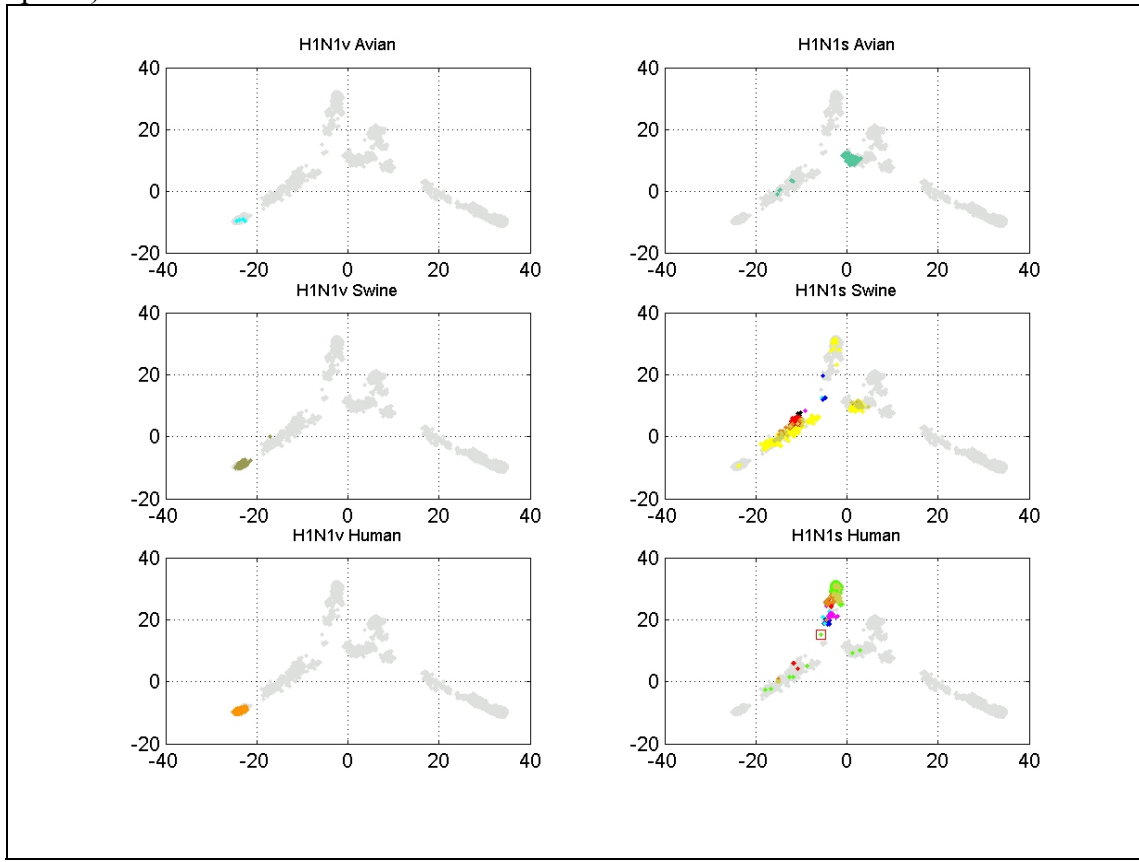

**Figure 1** – PC-1 *vs.* PC-2 scatter-plot of principal component analysis of the ISSCOR descriptors for the 9131 full-length hemagglutinin sequences (light gray points), with sequences corresponding to the 2009-2010 pandemic (H1N1v, left column), and the seasonal (H1N1s, right column) H1N1 serotypes, are color-coded according to their infected hosts. The early XX c. sequences isolated from the swine (in the middle-right panel) and the human (in the bottom-right panel) are color coded as follows: one case of "Spanish flu" (strain A/South_Carolina/1/1918) – red square; sequences from 1930' – blue; from 1940' – cyan; from 1950' – magenta; from 1960' – black; from 1970' – red; from 1980' – brown; from 1990' – khaki; and from 2000' and 2010' – light green (human) and yellow (swine) points.

Schafer *et al.* [35] determined the most probable avian origin of the Asian flu pandemic, and shown antigenicity of H2 HAs from representative human and avian viruses, as well as their evolutionary characteristics in respective hosts.

| accession | CY008988 | CY125862 | CY026283 | CY020381 | CY087800 | CY087792 |
|---|---|---|---|---|---|---|
| serotype | H1N1 (human) | H1N1 (human) | H1N1 (swine) | H2N2 (human) | H2N2 (human) | H2N2 (human) |
| strain | A/Denver/ 1957 | A/Kw/1/ 1957 (China) | A/swine/Wisconsin/ 1/1957 | A/Albany/26/ 1957 | A/Japan/ 305-MA12/ 1957 | A/Singapore/ 1-MA12E/ 1957 |
| A/Denver/1957 | 0 | 15 | 1012 | 1218 | 1220 | 1218 |
| A/Kw/1/1957 | 15 | 0 | 1012 | 1219 | 1221 | 1219 |
| A/swine/Wisconsin/1/ 1957 | 1012 | 1012 | 0 | 1218 | 1217 | 1215 |
| A/Albany/26/1957 | 1218 | 1219 | 1218 | 0 | 6 | 3 |
| A/Japan/ 305-MA12/1957 | 1220 | 1221 | 1217 | 6 | 0 | 7 |
| A/Singapore/1- MA12E/1957 | 1218 | 1219 | 1215 | 3 | 7 | 0 |

**Table II**

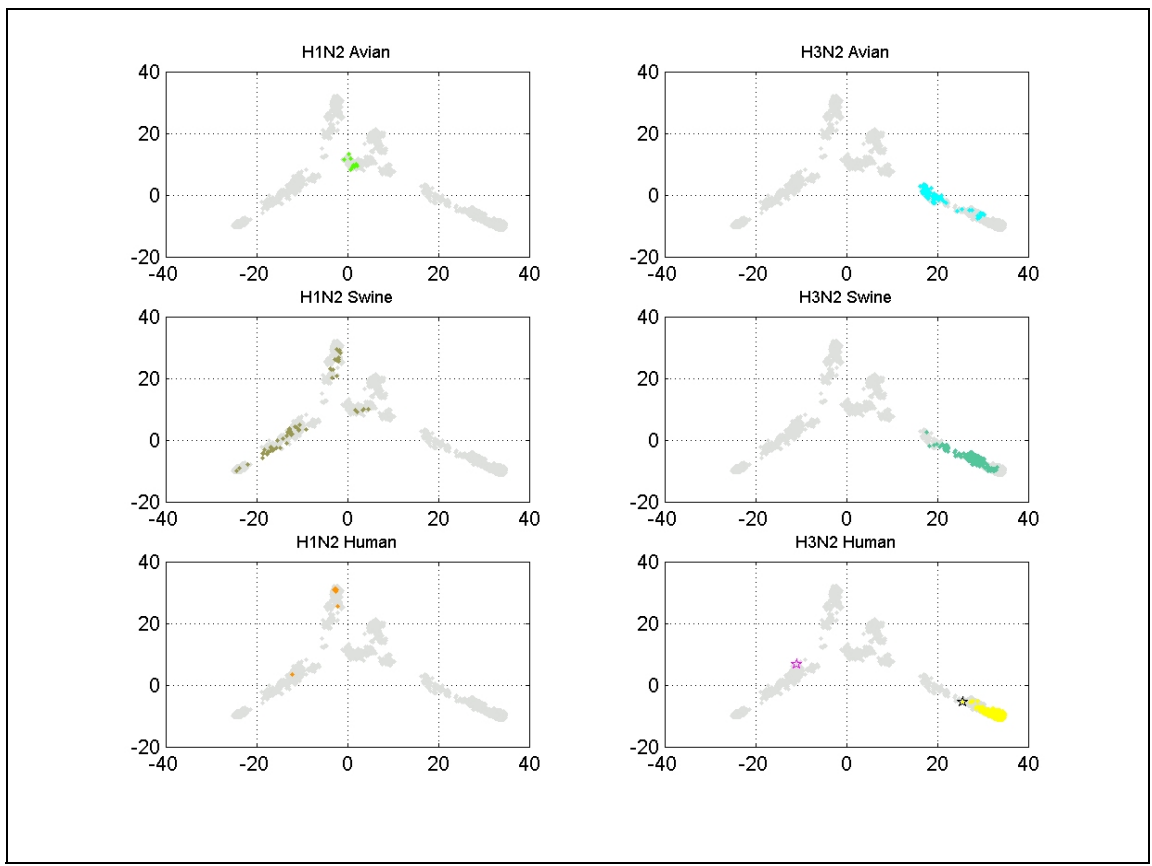

**Figure 2A** – PC-1 *vs.* PC-2 scatter-plot of principal component analysis of the ISSCOR descriptors for the 9131 full-length hemagglutinin sequences (light gray points), with sequences corresponding to the H1N2 (left column), and H3N2 (right column) serotypes, are color-coded according to their infected hosts. Additionally**: on the bottom-right panel** – the reference Hong Kong 1968 pandemic strains are marked: A/Hong_Kong/1-10-MA21-1/1968 (CY080523, black star), A/Hong_Kong/1-4-MA21-1/1968 (CY080515, black star), as well as the A/swine/Wisconsin/1/1968 (EU139825, of H1N1 subtype; magenta star).

Supposedly [35], the human H2 HAs, that circulated subsequently in the 1957-1968 period, formed a separate phylogenetic lineage, most closely related to the Eurasian avian H2 Has, and the antigenically conserved counterparts of the human Asian pandemic strains of 1957 might still continue to circulate in the avian reservoir, continuously coming into a close proximity with susceptible human populations. There was also an increased prevalence of H2 influenza viruses among wild ducks in 1988 in North America [35], preceding the appearance of H2N2 viruses in domestic fowl. As the prevalence of avian H2N2 influenza viruses increased on turkey farms and in live bird markets in New York City and elsewhere, greater numbers of these viruses have come into direct contact with susceptible humans. Unfortunately, the earliest avian host full HA sequence in our 9131 set was isolated only in the 1969, so we can't correlate their findings on the ISSCOR maps.

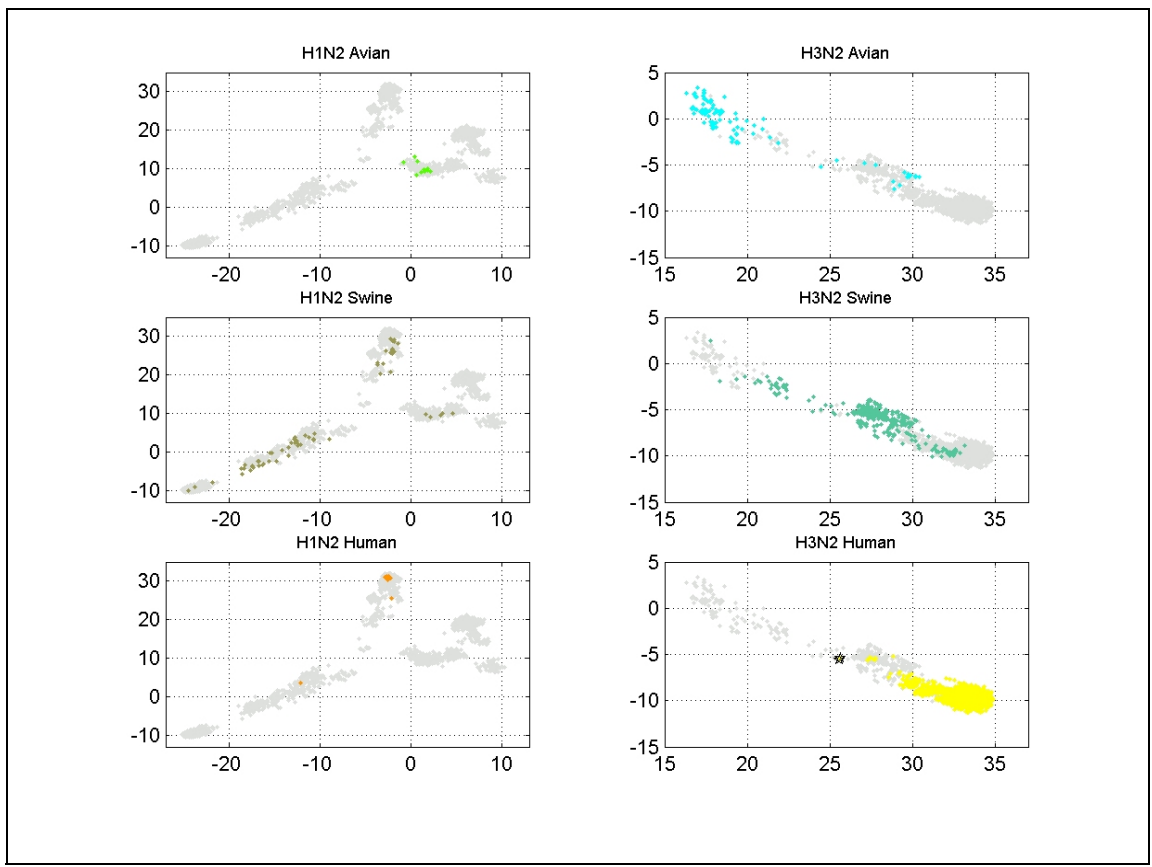

**Figure 2B** – PC-1 *vs.* PC-2 scatter-plot of principal component analysis of the ISSCOR descriptors for the 9131 full-length hemagglutinin sequences (light gray points), with sequences corresponding to the H1N2 (left column), and H3N2 (right column) serotypes displayed here are the same as on **Fig. 2A**, but now showing only the PC-1 and PC-2 regions where the respective H1N2 and H3N2 orthologs were present. Additionally**: on the bottom-right panel** – two reference Hong Kong 1968 pandemic strains are marked as black stars: A/Hong_Kong/1-10-MA21-1/1968 (CY080523), A/Hong_Kong/1-4-MA21-1/1968 (CY080515).

Then in July 1968 the next pandemic's virus H3N2 was first isolated in Hong Kong [36]. Again, there are only three strains isolated in 1968 among 9131 HAs strains: A/Hong_Kong/1-10-MA21-1/1968 (CY080523), A/Hong_Kong/1-4-MA21-1/1968 (CY080515), and A/swine/Wisconsin/1/1968 (EU139825, of H1N1 subtype) – the

first two are closely related (six mutations distant from each other), while the third one is distant from both by about 1200 nucleotide differences, and clearly it was not likely to be a pandemic precursor. Scholtissek *et al.* [37] concluded that the H3N2 subtype was presumably derived from a H2N2, by retaining seven segments of the H2N2, while the gene coding for the HA was recombined from Ukrainian duck or another highly related avian strain (not present on the ISSCOR map here).

*The swine-like virus 2009-2010 H1N1 pandemic*

In contrast to the previous outbreaks, during the 2009 "swine flu" H1N1 pandemic an abundant collection of sequential data was gathered in concert all across the globe. Interestingly, among the 178 pandemic strains there are two: the A/Malaysia/2142295/2009 (CY119330), and the A/Malaysia/2142299/2009 (CY119338), both collected on Jan. 9$^{th}$, that is much earlier than the commonly acknowledged "start" of the 2009 pandemic in March.

The **Figure S1** (*suppl. on-line materials*) shows the NJ+ phylogram of the 2009 pandemic H1N1 HA all early 178 sequences (isolated between Jan. 1$^{st}$ and April 30$^{th}$). The tree was calculated together with the 380 putative precursor sequences of other serotypes (all collected during the same period of Jan. 1$^{st}$ and April 30$^{th}$). The two most probable swine precursors: the H1N1 A/swine/Missouri/46519-5/2009 (HQ378741), and the H1N2 A/swine/Hong_Kong/NS252/ 2009 (CY085998), form a small sub-clade, adjacent to the newly emerged pandemic strains (which form together one tight cluster); with all the other possible ancestors markedly distant. The CY085998 one is closer to the pandemic genes by few mutations than the HQ378741 strain. The CY085998 is a triple reassortant swine strains, present in a large study of phylogenetic evolution relationships of putative precursors to the 2009 H1N1 pandemic, examined in great details in [38] – and although it wasn't indicated as a most probable human pandemic precursor by the authors, its position on their NJ tree (*c.f.* Fig. S2a in *suppl. materials* of [38]) marks it as a very likely candidate[1].

As the incubation and infectivity period of the virus lasts about one week, it is clear that the precursors ought to be extant at a time of possible genetic shift event to occur, however, we did repeat the analysis including also as an additional candidates all the 569 strains isolated during 2008. Partial results are shown on **Figure S2** (*suppl. materials*), besides two sequences described, there were also three other porcine H1N1 strains, preceding the HQ378741, in the same small sub-clade: the A/swine/North Carolina/3793/2008 (JQ624667), the A/swine/Illinois/02064/2008 (CY099095), and the A/swine/Ohio/02026/2008 (CY099159), confirming validity of the former analysis. The relationships between these five putative precursor sequences on the ISSCOR map are shown on **Figure S3** (*suppl. materials*).

The pandemic 2009 H1N1 cluster contains also five other porcine-host HA genes, but they are all more recent than May 2009: the H1N2 (circles) – A/swine/ Italy/116114/ (CY067662), A/swine/Minnesota/A01076209/2010 (JQ906868), and A/swine/Nebraska/ A01203626/2012 (JX444788); and the H1N1 (crosses) – A/swine/Illinois/ A01076179/2009 (JX042553, isolated Dec. 6$^{th}$, 2009), and

---

[1] Two other H1N2 porcine strains they have also studied in [38] – the A/swine/Hong_Kong/1435/2009 (CY061653, *c.f.* Fig. S3, square) was isolated only later, on July 23$^{rd}$ 2009, and the A/swine/Hong_Kong/2314/2009 (CY061741, not in our 9131 set) was isolated on Oct. 22$^{nd}$ 2009.

A/swine/Shepparton/6/2009 (JQ273542, isolated Aug.17[th], 2009). In line with our observations Garten *et al.* [39] have found that molecular markers predictive of adaptation to humans were not present in the early (as of May 2009) pandemic H1N1 viruses, and that antigenically the viruses were homogeneous – similar to North American swine H1N1 viruses, but distinct from seasonal human H1N1 isolates.

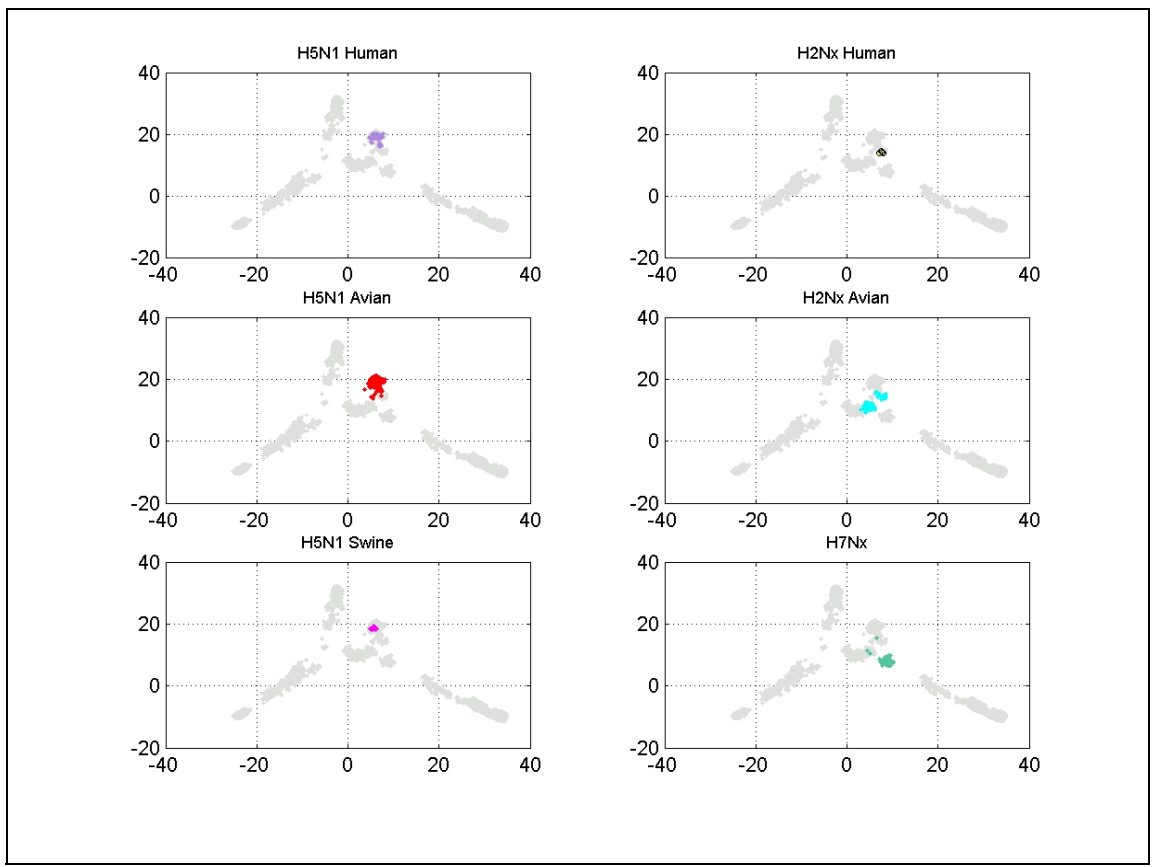

**Figure 3A** – PC-1 *vs.* PC-2 scatter-plot of principal component analysis of the ISSCOR descriptors for the 9131 full-length hemagglutinin sequences (light gray points), with sequences corresponding to the H5N1 (left column), H2Nx (right column, top and middle), and H7Nx (right column, bottom) serotypes.

Similarly, Smith *et al.* [40] concluded that the initial transmission to humans must have occurred several months before recognition of the outbreak Moreover, the unsampled history prior to pandemic means that the nature and location of the genetically closest swine viruses revealed little about the immediate origin of the epidemic.

Surprisingly if we compare the long dominance of over 40 years for the H3N2 serotype, already at two years since the 2009 pandemic outbreak, the new H1N1 variant displays now rather low abundance. Of the 190 full length HAs isolated during eleven months of 2012 till Dec. 9[th] there were – from porcine hosts: 115 of the H1N1 seasonal serotype, 2 of H1N2, 34 of the H3N2; from avian hosts: 10 of the H5N1, and 1 of the H7N3; from human hosts: 1 of the H7N3, 3 of the H5N1, 14 of the H3N2; however, there were only 10 of the recently dominant H1N1 2009-2010 pandemic type strains. It might be of a high interest to observe the behavior of swine-like H1N1 strains during the forthcoming 2012-2013 flu season, and its prevalence ratios in comparison especially to the H3N2 isolates.

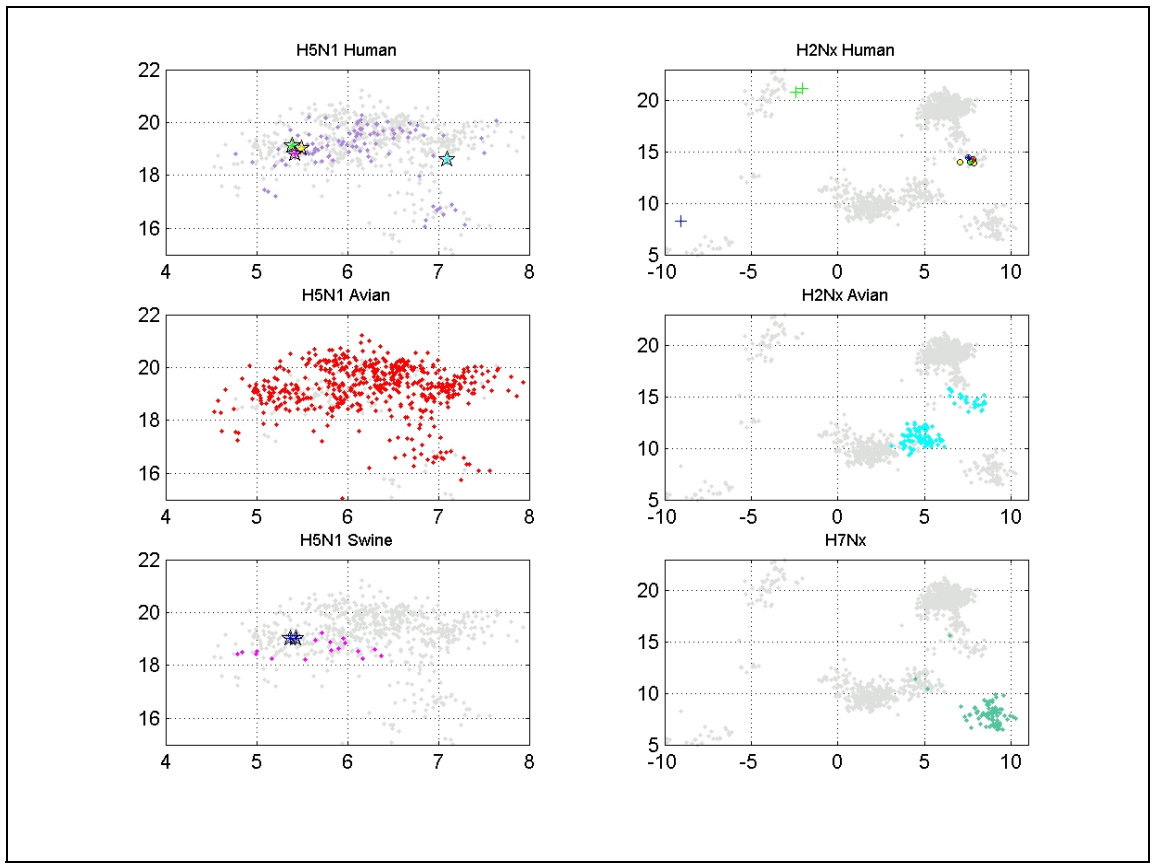

**Figure 3B** – PC-1 *vs.* PC-2 scatter-plot of principal component analysis of the ISSCOR descriptors for the 9131 full-length hemagglutinin sequences (light gray points), with sequences corresponding to the H5N1 (left column), H2Nx (right column, top and middle), and H7Nx (right column, bottom) serotypes – same as on **Fig. 3A**, but showing only the PC-1 and PC-2 regions where the respective H5N1 and H2N2 orthologs were present. Additionally**: on the top-left panel** – the reference wild type strains of the A/Indonesia/5/2005 isolated from human hosts are marked individually (accessions: CY116646 – yellow, the Fouchier's sequence from [24]; EU146622 – green; GQ149235 – magenta; and AY651334 – cyan, the /VietNam/1203/2004 Imai's strain [23]). The two blue stars in the bottom-left panel belong to the Fouchier's ferret sequences (CY116654 of their ferret #1, and CY116662 for ferrets #2 to #7); and **on the top-right panel** – the reference sequences of the 1957 Asian flu H2N2 (*c.f.* also **Table II**) are marked: CY087792 A/Singapore/1-MA12E/1957 – blue; CY087800 A/Japan/305-MA12/1957 – green; CY020381 A/Albany/26/1957 – red; as well as 1957 sequences of H1N1 serotype: CY008988 A/Denver/1957 and CY125862 A/Kw/1/1957 – green crosses (human), and CY026283 A/swine/Wisconsin/1/1957 (porcine).

*Laboratory-induced transition to the droplet-transmission infectivity*

The **Table ST1** (*supplementary on-line materials*) contains the list of accession numbers for all H5N1 sequences carrying the full pattern of the seven nucleotides (C355, C510, A514, A713, G718, A720, C992) required to be present in order to undergo the transition [24] from avian transmissible (the "before" state) to mammals airborne-transmissible (the "after" state, i.e. carrying the pattern: T355, A510, G514, T713, A718, C720, T992). The distributions of genes carrying each of the seven individual nucleotides, specific to this transition (*c.f.* [24] and their table S1 on page 15 of supplementary materials – pls. note that the numeration of nucleotides in their table differs from the shown here – our numbering starts always from the UAG codon of each gene) are mapped on **Figs. 4A** "before", and **4B** "after" transition. The distribution of the single-individual "after" positions is quite strikingly different from the distribution of the single-individual "before" positions – too numerous to list here, except for the T355 position (present only in 17 strains; top-left panel on **Fig. 4B**), and the T713 position (present only in 23 sequences; second top-right panel on **Fig. 4B**).

It is also noteworthy, that as this transition can take place only among H5N1 strains, as all other strains are already capable of droplet borne infectivity in mammals, the three mutations: C355T, A713T, and A720C are indeed crucial for the emerging strain – among all the 9131 alleles they are present together in just a single pair of sequences: CY116646 and CY116654, indicative of their artificial, laboratory origin. Indeed, the visualization which of the "before" seven mutated positions were present with just one exception (that is which genes were carrying six out of the seven "before" nucleotides, *c.f.* **Fig. 4S** in *suppl. materials*) is quite striking. In contrast to a very broad dispersal for the single "before" positions as seen on **Fig. 4A** – encompassing each of HA serotypes; all of the six-out-of-seven cases were present exclusively among H5N1 isolates (mostly avian, but also 8 porcine, and 54 human, *c.f.* **Table ST1**); ranging in numbers from 220 (positions: C355, G718, C992; and also the all seven positions) to 404 (for the position A514) sequences. Our results differ from these of Russell *et al.* [25] who performed a phylogenetic analysis of some H5N1 strains to reveal temporal, and to some extent also spatial, distributions of the between two to five of these seven mutations. On the one hand, it might seem that 220 orthologs, each carrying already all seven of the prerequisite "before" positions indicate quite substantial presence within all H5N1 strains in the whole data set. However, the comparison to just one sequence in the mutated "after" state clearly indicates rather very small probability of such an event occurrence.

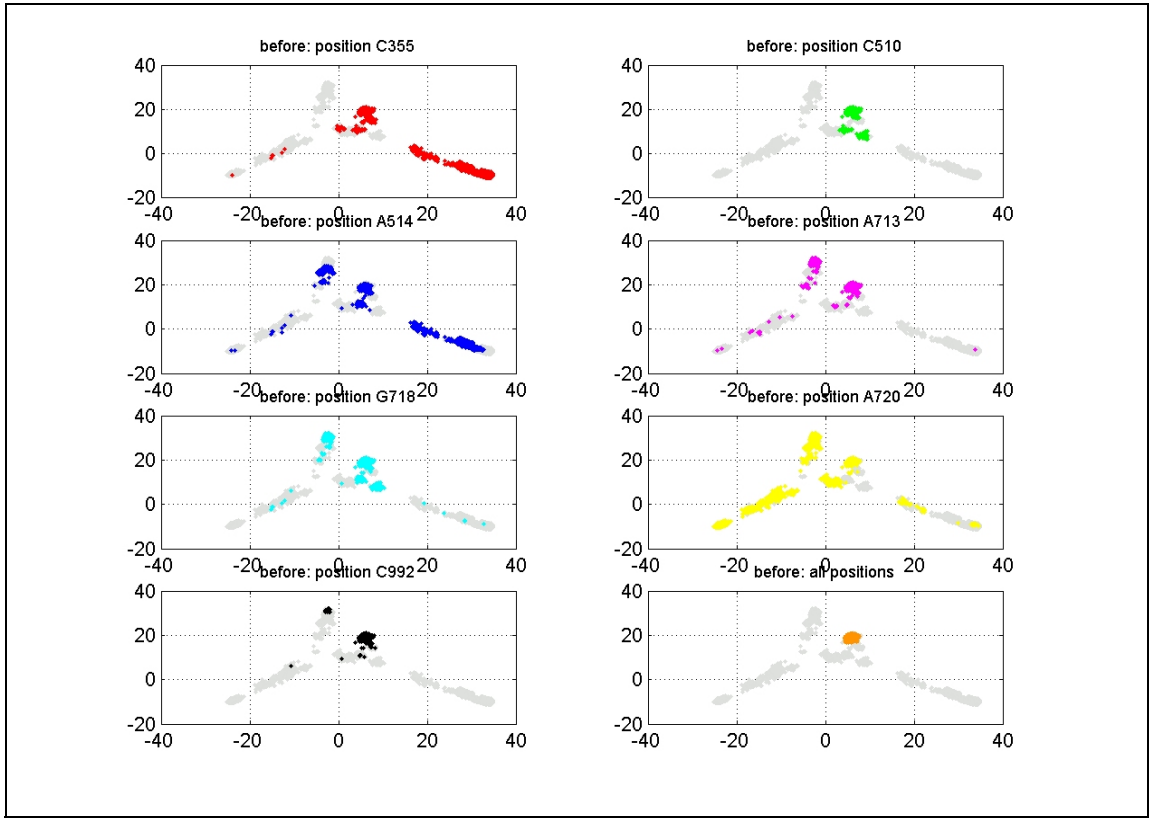

**Figure 4A** – the map of all sequences carrying the "before" transition individual nucleotide positions

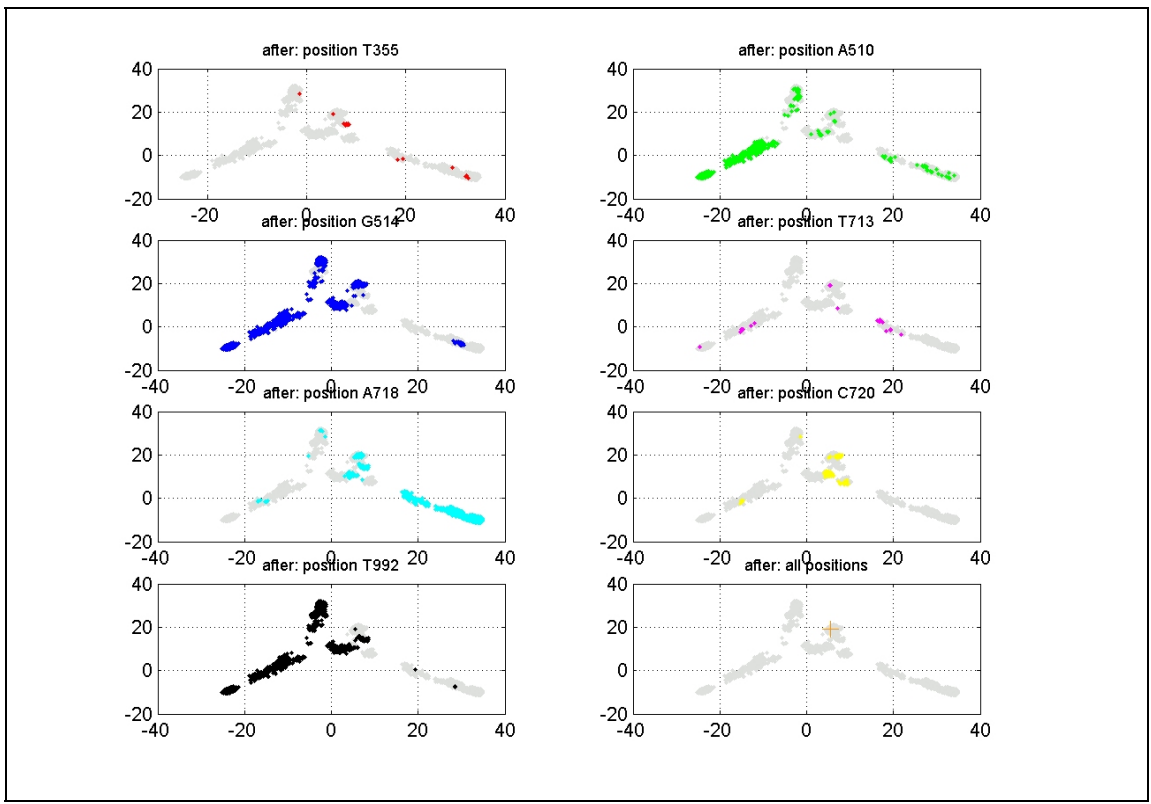

**Figure 4B** – the map of all sequences carrying the "after" transition individual nucleotide positions

## *4. Conclusions*

The ISSCOR-PCA method enables fast and efficient visualization of evolutionary relations present in a very large, complex set of homologous sequences – otherwise not an easy and rather tedious task for other phylogenetic analysis algorithms available, when it involves many thousands of sequences. Our approach significantly simplifies the effort, by producing two-dimensional projections from the multidimensional hyperspace of descriptors characterizing each of individual strains, allowing a clear evaluation of the genetic diversity ranging inside such large set of homologs.

Based on the size of the sequences 9131 set, we have examined the odds of putatively tracking an origins of all flu pandemics in XX c, however, their mostly unsampled history and especially severe paucity of zoonotic data for all major genetic shifts prior to the H1N1 pandemic of 2009-2010 deemed the task not possible. In contrast, for the 2009 "swine flu" H1N1 pandemic an abundant collection of sequential data was gathered from human hosts, but again not so many from the porcine or avian ones. The ISSCOR maps confirmed the close affinity of the earliest HAs of human isolates to their tentative precursors from swine, and yet even for this well documented epidemic the nature and location of the genetically closest swine viruses reveal little about the immediate origin of the infection. Clearly, much higher ratio of porcine and avian isolates needs to be routinely monitored in future for analyses to feasibly pinpoint inter-species acts of transmission, even if only in *ex post* descriptive a manner.

In an interesting study Renzette *et al.* [41] examined *ab initio* passaging of the A/Brisbane/59/2007 (H1N1), and the A/Brisbane/10/2007 (H3N2) in MCDK cell cultures, followed by a deep sequencing study, and demonstrated that some surprises might await there, as they have shown rather unexpected *increase* in both serotypes' viral diversity – occurring concurrently at the same time, albeit in two separate cell lines. As a deep sequencing *ab initio* experiments produce large amounts of reliable evolutionary data it would be of interest to study them in much more detail. Therefore, it is of high interest to apply the ISSCOR-PCA method presented here to such an *ab initio* dataset, optimally also in conjunction with ancestry tracking based on disentangled neighbor joining trees [26].

Tracing the chronology of individual strains isolation times on the maps specific to different serotypes revealed that oldest strains occupy mostly positions in the middle of the roughly triangular shape (*c.f.* **Fig. 1**) distribution, whereas newer strains spread gradually towards apexes of that triangle. This point is of importance – our analysis shows that the most distant sequences of hemagglutinin were all isolated from human hosts: A/Hamburg/1/2005 (H1N1 seasonal), A/Sydney/DD3_17/2010 (H1N1 pandemic 2009-2010); and A/Thailand/Siriraj-06/2002 (H3N2). The ISSCOR analysis, unexpectedly for us, showed that the hemagglutinin variability is largest in case of strains invading humans, and seems to be less pronounced in case of strains detected in birds. However, as sampling is much skewed towards human strains (*c.f.* **Table I**), the statistics of the collected set do not allow for any far–reaching hypotheses concerning species' specific virus–host interactions. Nevertheless, taking into account the sheer volume of the data analysed we propose that the edges of this assembly delimit the extent of genetic diversification of the influenza virus hemagglutinin. This bears on the immunological variability of this gene, allowing for broader look on influenza epidemiology.

## *Conflicts of interests*

Authors declare no conflict of interests.

## *Acknowledgements*

We would like to thank Pat Churchland for looking over the English. This work was partially supported by the EU project SSPE-CT-2006-44405, and also partially supported from the BST/115/30/E-343/S/2012.

Additional partial funding was generously provided by the WND-POIG.01.01.02-00-007/08 grant from the European Regional Development Fund.

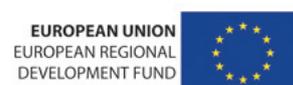

## *Supplementary on-line materials*

**Table ST1** contains the list of accession numbers for all H5N1 sequences carrying full configuration of nucleotides (C355, C510, A514, A713, G718, A720, C992) required to undergo transition from avian transmissible to mammals droplet-transmissible (i.e. carrying the configuration: T355, A510, G514, T713, A718, C720, T992).

| host | number | accessions |
|---|---|---|
| avian | 158 | EU124091, EU124082, EU124094, EU124158, EU124157, EU124093, EU124083, EU124153, EU124163, EU124162, EU124160, EU124155, EU124151, EU124149, EU930980, EU931004, EU124159, FJ784842, CY091811, CY091812, CY091815, CY091816, CY091819, CY091820, CY091822, CY091797, CY091799, CY091800, CY091801, CY091802, CY091803, CY091874, CY091878, CY091887, CY091886, CY091789, CY091774, CY091776, CY091777, CY091778, CY091779, CY091782, CY091783, CY091784, CY041290, GQ184236, GQ184238, GQ184239, JN807782, JN807777, JN807803, JN807785, JQ858472, JN807840, JN807843, CY062601, FJ686831, CY062602, FJ686832, CY062603, CY062605, CY062606, CY062607, CY062608, FJ686833, CY062609, FJ686836, CY062610, JF357723, CY099579, CY091788, CY091787, CY091785, CY091786, FJ784843, JQ809274, EF670482, FJ784851, FJ784853, FJ784847, FJ784854, HM466695, CY091804, CY091823, CY091806, CY091807, CY091809, CY091810, CY091895, CY091913, CY091942, CY091943, CY091944, HM208702, CY091905, CY091906, CY091907, CY091908, CY091915, JF827081, JF827078, CY091955, CY091889, CY091901, CY091910, CY091966, CY091795, CY091793, CY091794, CY047457, EU497921, EF541406, CY095680, CY095686, CY095701, JN055363, EU930892, EU124276, EU124277, CY091951, CY091859, CY091860, CY091870, CY091946, GU727669, FJ784844, FJ784845, FJ784846, FJ784848, FJ784849, FJ784850, FJ784855, FJ784856, FJ784841, CY091897, CY091953, CY091867, CY091949, CY091862, CY091865, CY091893, CY095683, CY095692, CY095695, CY095698, CY095704, EU930940, JX021306, JX021305, GU220793, HM627945, HM627913, HM627921, HM627929, HM627937, JF732739, CY091781, JN582041 |
| human | 54 | CY014280, CY014465, CY014489, CY014497, CY014529, CY017638, CY017654, CY017662, CY017670, CY017688, CY019376, CY098603, CY098634, CY098641, CY098655, CY098668, CY098681, CY098688, CY098695, CY098702, CY098716, CY098730, CY098737, CY098751, DQ371928, DQ371930, EU146622, EU146640, EU146648, EU146681, EU146688, EU146755, EU146777, FJ492879, FJ492880, FJ492881, FJ492882, FJ492883, FJ492884, FJ492885, FJ492886, GQ466176, HM114545, HM114561, HM114569, HM114577, HM114585, HM114593, HM114609, HM114617, AY651334, AY818135, CY116646, GQ149235 |
| swine | 8 | AY646424, DQ997253, DQ997262, HM440083, HM440091, HM440123, HM440131, HM440147 |

**Table ST1**

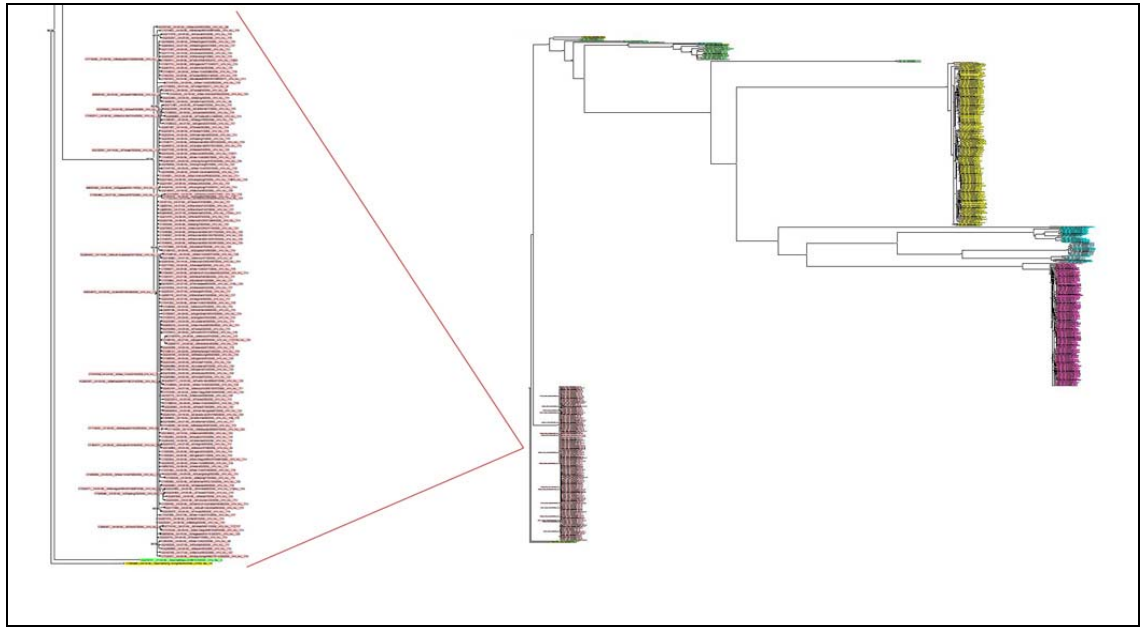

**Figure S1** – The NJ+ phylogram of the 2009 pandemic H1N1 hemagglutinin early 178 sequences (light rose, and the enlarged inset), which were isolated between Jan. 1st and April 30th. Calculated together with 380 putative precursor sequences of other serotypes (collected during the same period): seasonal H1N1 (yellow – human hosts, and light green – porcine hosts), H3N2 (magenta – human), H5N1 (blue – avian; and light blue – human), H1N2 (olive – porcine). The zoomable Fig_S1_large.pdf file can be used for a better visibility of details. The two most probable swine precursor gene sequences are located at the very bottom on the left: the H1N1 A/swine/Missouri/46519-5/2009 (HQ378741, light green) and the H1N2 A/swine/Hong_Kong/NS252/2009 (CY085998, olive).

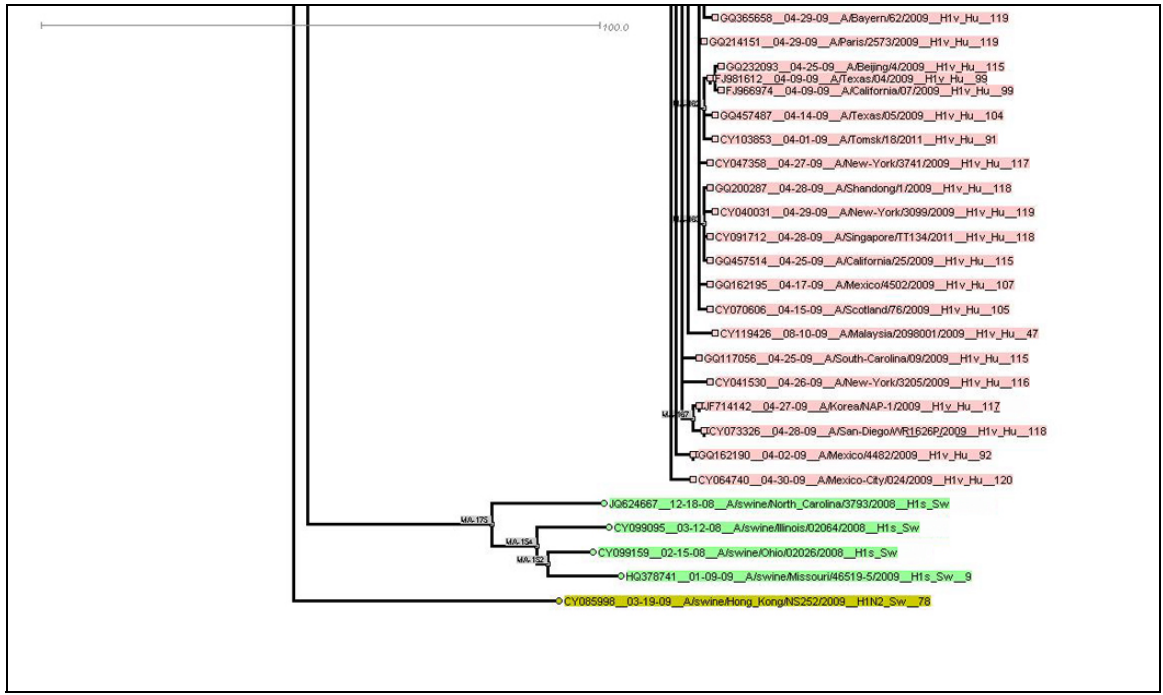

**Figure S2** – The NJ+ phylogram of the 2009 pandemic H1N1 hemagglutinin early 178 sequences (light rose), which were isolated between Jan. 1st and April 30th. Calculated as that on **Fig. S1**, but besides the 380 putative precursor sequences of other serotypes, collected during the same period, also additional 569 candidate genes, collected during the whole 2008 year were used. The result was that in addition to the two most probable swine precursor gene sequences are located at the bottom: the H1N1 A/swine/Missouri/46519-5/2009 (HQ378741, light green) and the H1N2 A/swine/Hong_Kong/NS252/ 2009 (CY085998, olive), also three other porcine H1N1 strains (light green), preceding in the same, small subclade are present: the A/swine/North_Carolina/3793/2008 (JQ624667), the A/swine/Illinois/ 02064/2008 (CY099095), and the A/swine/Ohio/02026/2008 (CY099159).

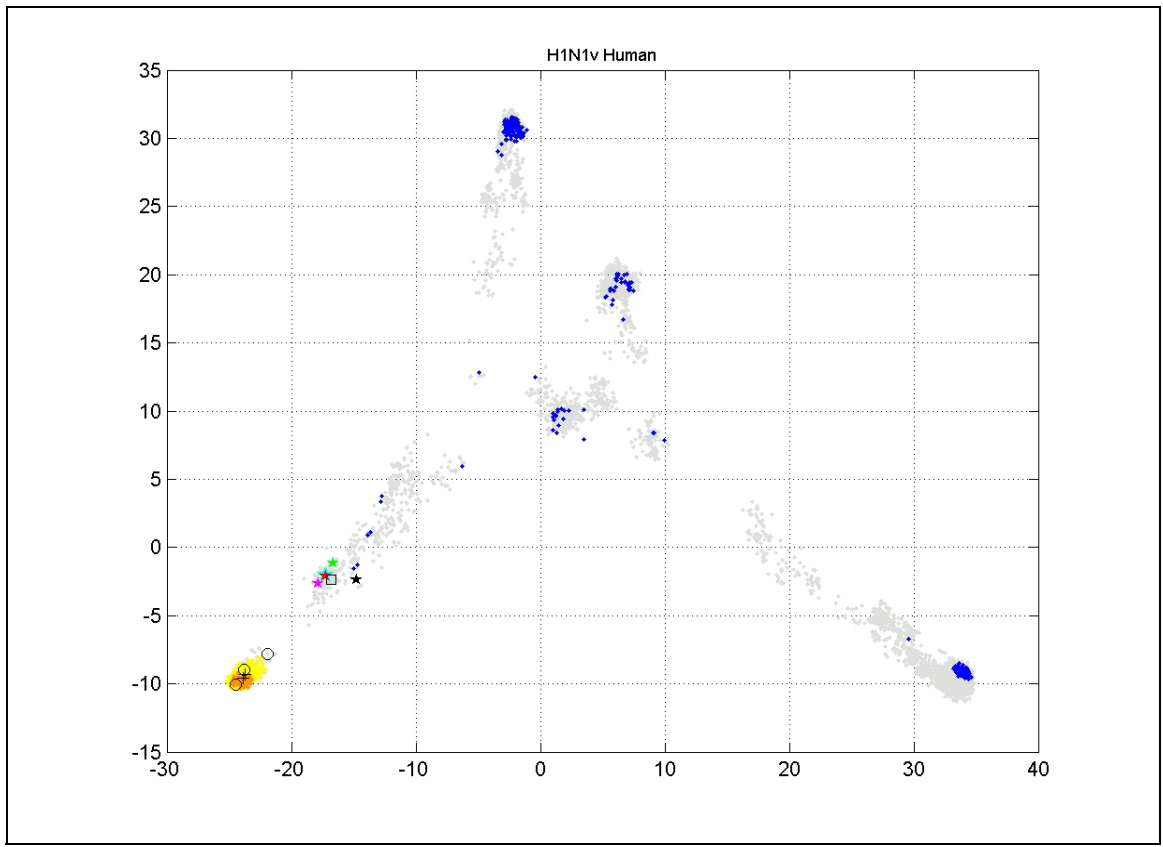

**Figure S3** – PC-1 *vs.* PC-2 scatter-plot of principal component analysis of the ISSCOR descriptors for the 9131 full-length hemagglutinin sequences (light gray points), with the sequences corresponding to the hemagglutinin early 178 sequences (isolated between Jan. 1st and April 30th, orange points; *c.f.* the bottom-left panel in **Fig. 1**; the rest of pandemic human isolates is marked yellow). The 380 putative precursor sequences of other serotypes, collected during the same period are also mapped (blue). The two most probable swine precursor gene sequences (from the phylogram on **Fig. S1**) are: the H1N1 A/swine/Missouri/46519-5/2009 (HQ378741, red) and the H1N2 A/swine/Hong_Kong/NS252/ 2009 (CY085998, black). Also three other porcine H1N1 strains (as in the legend to **Fig. S2**) are shown: the A/swine/North_Carolina/3793/2008 (JQ624667, cyan), the A/swine/Illinois/ 02064/2008 (CY099095, light green), and the A/swine/Ohio/02026/2008 (CY099159, magenta). Yet another H1N2 swine strain – A/swine/Hong Kong/1435/2009 (CY061653, square) was isolated later, on July 23rd.

Within the pandemic 2009 H1N1 cluster (at approx. PC-1 = -24, and PC-2 = -10) there are also five other porcine-host HA genes, but they are all more recent than May 2009: the H1N2 (circles) – A/swine/ Italy/116114/ (CY067662), A/swine/Minnesota/A01076209/2010 (JQ906868), and A/swine/Nebraska/ A01203626/2012 (JX444788); and the H1N1 (crosses) – A/swine/Illinois/ A01076179/2009 (JX042553, isolated Dec. 6th, 2009), and A/swine/Shepparton/6/2009 (JQ273542, isolated Aug.17th, 2009). It is possible that in those cases infection might have followed a reversed human → swine paths.

It is interesting to note that while all human strains of the pandemic 2009 H1N1 isolates are mapped (orange points on the **Fig. S3**) at the very tip of their cluster, the later strains of the 2009 pandemic (yellow points) map more towards the "common origin" location. Similarly like it was the case of H1N1 strains of the 1918, 1930', 1940', 1950', etc. clusters (*c.f.* bottom-right panel of the **Fig. 1**) described in the main text.

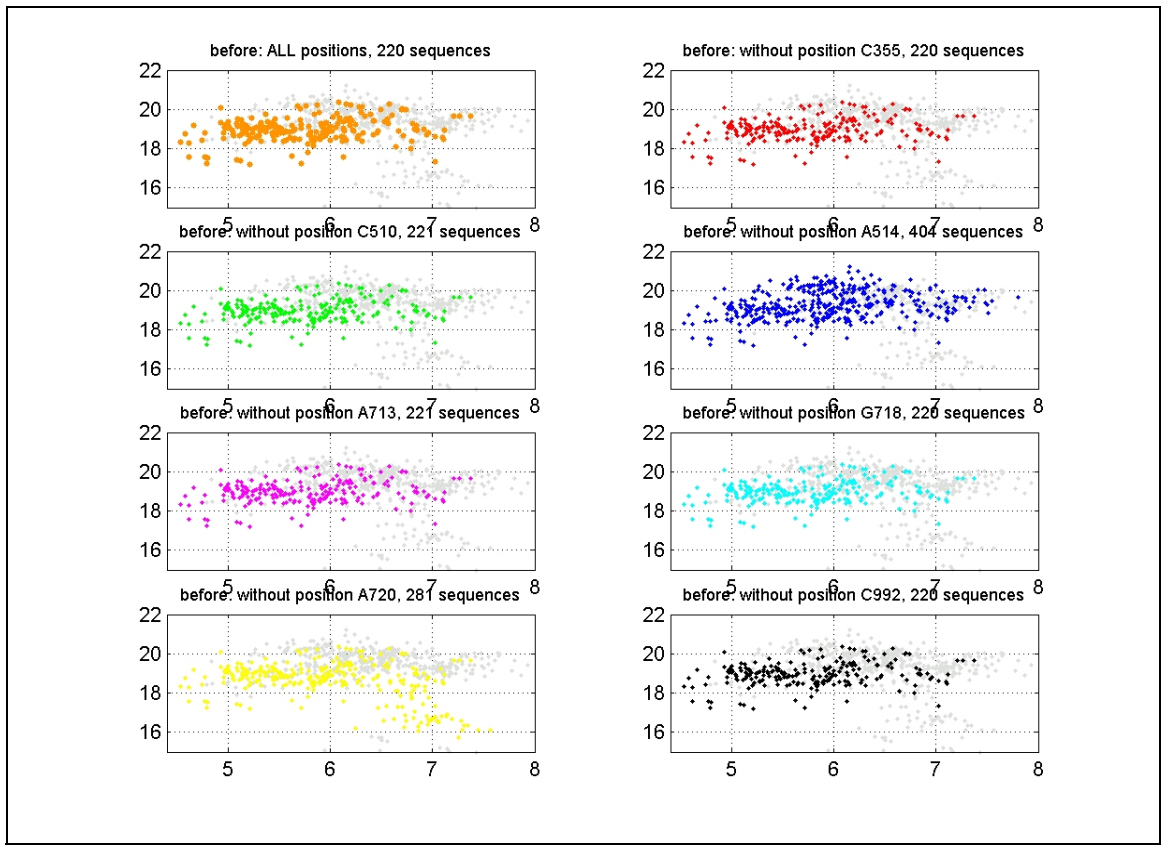

**Figure S4** – The enlarged fragment of the PC-1 *vs.* PC-2 map ot of principal component analysis of the ISSCOR descriptors for the 9131 full-length hemagglutinin sequences (light gray points), with the sequences corresponding to the region of the H5N1 for the "before" state of avian transmissible to mammals airborne-transmissible. The sequences carrying all seven positions (as described in [24]), as well as carrying combinations of the six positions with exclusion the one indicated, are color-coded as shown.

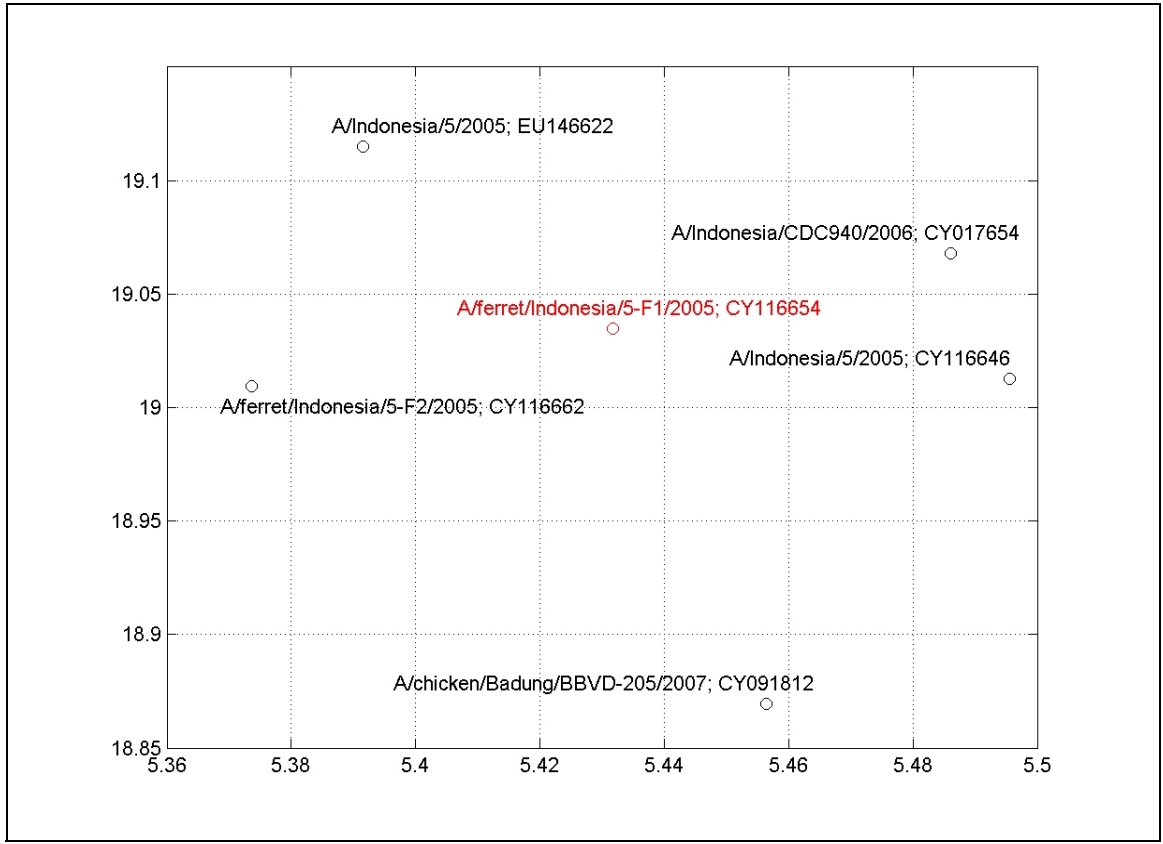

**Figure S5** – The fragment of the PC-1 *vs.* PC-2 scatter-plot of the principal component analysis of the ISSCOR descriptors for the hemagglutinin sequences map – the five strains that are closest to the droplet-infection transmissible ferret #1 of Herfst *et al.* [24]: A/ferret/Indonesia/5-F1/2005, CY116654 strain (*c.f.* **Figs. 4A**, **4B**, and **S4**).